\documentclass[aps,prb,twocolumn,showpacs]{revtex4}
\usepackage{bm}
\usepackage{epsfig}
\usepackage{graphicx}
\begin{document}

\title{Spin relaxation of conduction electrons in (110)-grown quantum wells}

\author{S.\,A.\,Tarasenko}

\affiliation{A.\,F.\,Ioffe Physical-Technical Institute, Russian
Academy of Sciences, 194021 St.~Petersburg, Russia}

\begin{abstract}
The theory of spin relaxation of conduction electrons is developed for zinc-blende-type quantum wells grown on (110)-oriented substrate. It is shown that, in asymmetric structures, the relaxation of electron spin initially oriented along the growth direction is characterized by two different lifetimes and leads to the appearance of an in-plane spin component. The magnitude and sign of the in-plane component are determined by the structure inversion asymmetry of the quantum well and can be tuned by the gate voltage. In an external magnetic field, the interplay of cyclotron motion of carriers and the Larmor precession of electron spin can result in a nonmonotonic dependence of the spin density on the magnetic field.
\end{abstract} 



\maketitle

\section{Introduction}

The long and tunable spin lifetime of carriers in semiconductor low-dimensional structures is a crucial factor for spintronic applications. Of particular interest in this context are quantum wells (QWs) grown from zinc-blende-type semiconductors on (110)-oriented substrates.
In such structures, the spin lifetime of conduction electrons can be as long as several nanoseconds at room temperature~\cite{Ohno99,Karimov03,Dohrmann04} and tens of nanoseconds at low temperature~\cite{Muller08} allowing for a long-range spin transport.~\cite{Couto07} Moreover, the spin lifetime can be widely tuned by the gate voltage~\cite{Karimov03,Hall05,Eldridge08} and modifying the doping profile.~\cite{Belkov08} These features are attributed to suppression of the D'yakonov-Perel' (DP) spin relaxation mechanism in symmetric (110) QWs.~\cite{Dyakonov86} The DP mechanism (Ref.~[\onlinecite{DP}]) usually limiting the electron spin lifetime is based on the precession of electron spins in an effective magnetic field induced by spin-orbit interaction in noncentrosymmetric media.
In (110)-grown QWs, the effective magnetic field caused by bulk inversion asymmetry (BIA) points along the growth direction.~\cite{Dyakonov86} Therefore, electron spins oriented along the QW normal do not precess in the field and the DP mechanism gets ineffective. If, however, the QW is asymmetric, structure inversion asymmetry (SIA) leads to the Rashba effective magnetic field~\cite{Bychkov84} which lies in the QW plane and speeds up the spin dephasing. Thus, by measuring the spin lifetime of carriers one can conclude on the Rashba field strength.

Here, we develop a microscopic theory of the electron spin relaxation in low-symmetry two-dimensional structures, such as QWs grown on (110)- and (113)-oriented substrates. 
We show that the mentioned above common analysis of spin dynamics
can only be used to illustrate the increase of spin lifetime in symmetric (110) QWs. However, the simplified model misleads in describing the spin dephasing in asymmetric structures and determining the Rashba field. In low-symmetry QWs, the growth direction is not a principle axis of the spin relaxation rate tensor. Therefore, the relaxation of electron spin initially oriented along the QW normal is described by a sum of exponential functions with different decay rates and leads to the appearance of an in-plane spin component.
The magnitude and sign of the in-plane component are determined by structure inversion asymmetry of the QW and can be tuned by the gate voltage. We also study the effect of an external magnetic field on spin relaxation and show that the interplay of the cyclotron motion of carriers and the Larmor precession results in a nonmonotonic dependence of the electron spin on the magnetic field.

It is worth noting that the coupling of the in-plane and out-of-plane components of electron spin considered here is a feature of structures of low point-group symmetry and does not occur in (001)-grown (even asymmetric) QWs. In (001)-oriented structures, both the Rashba and Dresselhaus fields lie in the QW plane and their constructive or destructive interference
leads to the in-plane anisotropy of spin relaxation.~\cite{Averkiev99,Kainz03,Averkiev06,Stich07} By contrast, in (110) QWs the Rashba and Dresselhaus fields are orthogonal and cannot compensate nor strengthen each other.

\section{Symmetry analysis}

Let us first consider the case of zero magnetic field. The time evolution of the spin density $\bm{S}(t)$, provided that the spin lifetime is longer than the carrier thermalization time, is described by~\cite{Dyakonov86}   
\begin{equation}\label{S_zeroB}
\frac{d S_{\alpha}(t)}{dt} = G_{\alpha} - \sum_{\beta} \Gamma_{\alpha\beta}
S_{\beta}(t)  \:,
\end{equation}
where $\bm{G}$ is the spin generate rate, e.g, due to optical pumping with circularly polarized light, $\Gamma_{\alpha\beta}$ are components of the spin relaxation rate tensor $\bm{\Gamma}$, and $\alpha$, $\beta$ are the Cartesian coordinates. The form of the tensor $\bm{\Gamma}$ depends on the spin relaxation mechanism and the QW point-group symmetry.  

Asymmetric quantum wells grown on (110)-oriented substrate as well as (113)-grown QWs are described by the point group $C_s$ which contains only two symmetry elements: identity and the mirror plane perpendicular to the QW plane.~\cite{Belkov08,Ivchenko_book} In the case of (110)-oriented structures, the mirror plane is normal to the in-plane axis $x\parallel[1\bar{1}0]$ and contains the axes $y\parallel[00\bar{1}]$ and $z\parallel[110]$. The symmetry analysis shows that non-zero components of the spin relaxation rate tensor in such QWs are $\Gamma_{xx}$, $\Gamma_{yy}$, $\Gamma_{zz}$, and $\Gamma_{yz}=\Gamma_{zy}$.

\begin{figure}[t]
 \centering\includegraphics[width=0.4\textwidth]{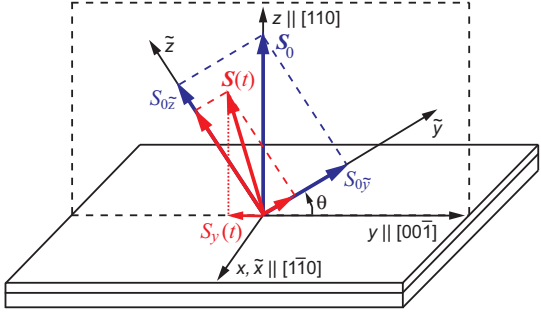}
 \caption{(Color online) Arrangement of the principle axes $\tilde{x}$, $\tilde{y}$, and $\tilde{z}$ of the spin relaxation rate tensor. Due to difference in the relaxation rates along the axes $\tilde{y}$ and $\tilde{z}$, the relaxation of electron spin $\bm{S}_0 \parallel z$ leads to the appearance of the in-plane component $S_y(t)$.}
 \label{figure1}
\end{figure}

The presence of the off-diagonal components $\Gamma_{yz}$ and $\Gamma_{zy}$ indicates that both the in-plane axis $y$ and the QW normal $z$ are not principle axes of the spin relaxation rate tensor. Therefore, the decay of $S_y$ (as well as $S_z$) cannot be described by a single spin lifetime. The principle axes $\tilde{x}$, $\tilde{y}$, $\tilde{z}$ and eigen values $\gamma_{i}$ of the tensor $\bm{\Gamma}$, i.e., relaxation rates along these axes, are found from the determinant
\begin{equation}\label{S_det}
\det (\bm{\Gamma} - \gamma \bm{I}) = 0 \:,
\end{equation}
where $\bm{I}$ is the unit matrix $3\times3$. Solution of Eq.~(\ref{S_det}) has the form
\begin{eqnarray}
\gamma_{\tilde{x}} &=& \Gamma_{xx} \:, \\
\gamma_{\tilde{y}} &=& \left[\Gamma_{yy}+\Gamma_{zz} + \sqrt{(\Gamma_{yy}-\Gamma_{zz})^2 + 4 \Gamma_{yz}^2} \right]/2 \:, \nonumber \\
\gamma_{\tilde{z}} &=& \left[\Gamma_{yy}+\Gamma_{zz} - \sqrt{(\Gamma_{yy}-\Gamma_{zz})^2 + 4 \Gamma_{yz}^2} \right]/2 \:, \nonumber 
\end{eqnarray}
where the axes $\tilde{x}$, $\tilde{y}$, $\tilde{z}$ are obtained from $x$, $y$, $z$ by rotating the coordinate frame around the axis $x$ by an angle $\theta$, see Fig.~\ref{figure1}. 
The angle $\theta$ is given by
\begin{equation}\label{theta}
\tan \theta = \frac{2\Gamma_{yz}}{\Gamma_{yy}-\Gamma_{zz}+\sqrt{(\Gamma_{yy}-\Gamma_{zz})^2+4\Gamma_{yz}^2}} \:.
\end{equation}

Solution of Eq.~(\ref{S_zeroB}) in the coordinate frame $(x,y,z)$ can be then derived by projecting $\bm{S}(t)$ and $\bm{G}$ onto the axes $\tilde{x}$, $\tilde{y}$, and $\tilde{z}$. This calculation shows that, for pulse excitation, the time evolution of electron spin $\bm{S}(t)$ after the pulse has the form
\begin{eqnarray}\label{S_time}
S_{x}(t) &=& S_{0x} {\rm e}^{-\gamma_{\tilde{x}}t} \:, \\
S_{y}(t) &=& S_{0y} (\cos^2{\theta} \, {\rm e}^{-\gamma_{\tilde{y}}t} + \sin^2\theta \, {\rm e}^{-\gamma_{\tilde{z}}t} ) \nonumber \\
&+& S_{0z} \cos\theta \sin\theta \, ({\rm e}^{-\gamma_{\tilde{y}}t} - {\rm e}^{-\gamma_{\tilde{z}}t}) \:, \nonumber \\
S_{z}(t) &=& S_{0z} (\cos^2{\theta} \, {\rm e}^{-\gamma_{\tilde{z}}t} + \sin^2\theta \, {\rm e}^{-\gamma_{\tilde{y}}t} ) \nonumber \\
&+& S_{0y} \cos\theta \sin\theta ({\rm e}^{-\gamma_{\tilde{y}}t} - {\rm e}^{-\gamma_{\tilde{z}}t}) \:, \nonumber
\end{eqnarray}
where $\bm{S}_0$ is the spin density at $t=0$. One can see that the relaxation of electron spin initially directed along the QW normal, i.e., when $\bm{S}_0 \parallel z$, is described by two different rates $\gamma_{\tilde{z}}$ and $\gamma_{\tilde{y}}$. Moreover, the relaxation leads to the rotation of spin in the $(yz)$ plane resulting in a non-zero value of $S_y$. Such a rotation is illustrated in Fig.~\ref{figure1} where the electron spin $\bm{S}(t)$ is shown as a sum of the components $S_{\tilde{y}}(t)$ and $S_{\tilde{z}}(t)$ along the principle axes $\tilde{y}$ and $\tilde{z}$.
At $t=0$, the spin is oriented along the QW normal, therefore, $S_{0\tilde{y}}=S_{0} \sin\theta$ and $S_{0\tilde{z}}=S_{0} \cos\theta$. In the course of time,
the components $S_{\tilde{y}}(t)$ and $S_{\tilde{z}}(t)$ decay at different rates, $\gamma_{\tilde{y}}$ and $\gamma_{\tilde{z}}$, respectively. Thus, at $t>0$, the ratio $S_{\tilde{y}}(t)/S_{\tilde{z}}(t)$ is not equal to the initial one $S_{0\tilde{y}}/S_{0\tilde{z}}$. It means that the spin $\bm{S}(t)$ does not point along the QW normal anymore and has non-zero component $S_y(t)$ as shown in Fig.~\ref{figure1}. Similarly, the relaxation of electron spin oriented along the $y$ axis leads to the appearance of out-of-plane component $S_z(t)$.   

Besides experiments with time resolution, spin phenomena are widely studied in the regime of
continuous-wave (cw) pumping, where the spin generation rate $\bm{G}$ is constant on the spin lifetime scale. The steady-state spin density in the regime of cw pumping can be directly obtained from Eq.~(\ref{S_zeroB}) and has the form 
\begin{eqnarray}\label{S_cw}
S_{x} &=& \frac{G_x}{\Gamma_{xx}} \:, \\
S_{y} &=& \frac{G_y \Gamma_{zz} - G_z \Gamma_{yz}}{\Gamma_{yy}\Gamma_{zz}-\Gamma_{yz}^2} \:, \nonumber \\
S_{z} &=& \frac{G_z \Gamma_{yy} - G_y \Gamma_{yz}}{\Gamma_{yy}\Gamma_{zz}-\Gamma_{yz}^2} \:. \nonumber
\end{eqnarray}
Equations~(\ref{S_time}) and~(\ref{S_cw}) are general and describe the spin properties of (110) QWs no matter which microscopic mechanism determines the spin dephasing. 

\section{D'yakonov-Perel' mechanism}

In a wide range of temperature, carrier density and mobility, the spin lifetime in two-dimensional semiconductor structures is limited by the DP mechanism.~\cite{DP} In this mechanism, components of the spin relaxation rate tensor in the collision-dominated regime are given by~\cite{Dyakonov86}
\begin{equation}\label{T_gen}
\Gamma_{\alpha\beta} =  - \int_{0}^{\infty} \frac{\tau_1}{f(0)} \frac{d
f(\varepsilon_{\bm{k}})}{d\varepsilon_{\bm{k}}}  \left[ \langle \bm{\Omega}_{\bm{k}}^2 \rangle \, \delta_{\alpha\beta} - \langle \Omega_{\bm{k},\alpha} \,\Omega_{\bm{k},\beta} \rangle \,\right] d\varepsilon_{\bm{k}} \:,
\end{equation}
where $\tau_1$ is the isotropization time of spin density, $f(\varepsilon_{\bm{k}})$ is the distribution function of carriers, $\varepsilon_{\bm{k}}=\hbar^2\bm{k}^2/(2m^*)$, $m^*$ is the effective mass, $\bm{\Omega}_{\bm{k}}$ is the Larmor frequency corresponding to the effective magnetic field caused by spin-orbit coupling, $\delta_{\alpha\beta}$ is the Kronecker symbol,  and the angle brackets denote averaging over the direction of the wave vector $\bm{k}$. The Larmor frequency corresponding to $\bm{k}$-linear effective magnetic field in (110) QWs of the $C_s$ point group has the form
\begin{equation}\label{Omega_k}
\bm{\Omega}_{\bm{k}} = \frac{2}{\hbar} (\alpha_1 k_y , -\alpha_2 k_x, \beta k_x) \:.
\end{equation}
The parameter $\beta$ originates from bulk inversion asymmetry while $\alpha_1$ and $\alpha_2$ are non-zero only in QWs with structure inversion asymmetry. We also note that $\alpha_1$ and $\alpha_2$ are linearly independent in QWs of the $C_s$ point group. The difference between $\alpha_1$ and $\alpha_2$ cannot be obtained in framework of the Rashba model which gives $\alpha_1=\alpha_2$. To obtain the difference in a microscopic calculation of the band structure one needs to take into account both QW asymmetry and the lack of an inversion center in the host crystal.~\cite{Cartoixa06} 

It follows from Eqs.~(\ref{T_gen}) and~(\ref{Omega_k}) that the components $\Gamma_{\alpha\beta}$ in asymmetric (110)-grown QWs assume the form
\begin{eqnarray}\label{Spin_time}
\Gamma_{xx} = (\alpha_2^2 + \beta^2) \,C \:, \;\; \Gamma_{yy} = (\alpha_1^2 + \beta^2) \,C \:, \\
\Gamma_{zz} = (\alpha_1^2 + \alpha_2^2) \,C \:, \;\; \Gamma_{yz} = \Gamma_{zy} = \alpha_2 \beta \, C \:, \nonumber
\end{eqnarray}
where the parameter $C$ depends on the temperature and carrier density, and for the degenerate two-dimensional electron gas it is given by $C =(4 \tau_1 m^*/\hbar^4) E_F$ with $E_F$ being the Fermi energy. Thus, for the DP mechanism we derive $\gamma_{\tilde{x}}= (\alpha_2^2 + \beta^2)C$, $\gamma_{\tilde{y}}=(\alpha_1^2+\alpha_2^2+\beta^2)C$, $\gamma_{\tilde{z}}= \alpha_1^2 C$, and $\tan \theta = \alpha_2/\beta$ (see Ref.~[\onlinecite{Cartoixa05}] for the case of $\alpha_1=\alpha_2$) . One can see that all three spin relaxation rates are different and the rate $\gamma_{\tilde{z}}$ differs from $\Gamma_{zz}$ approximately by a factor of 2 even in the case of small angles $\theta$.
\begin{figure}[t]
 \centering\includegraphics[width=0.49\textwidth]{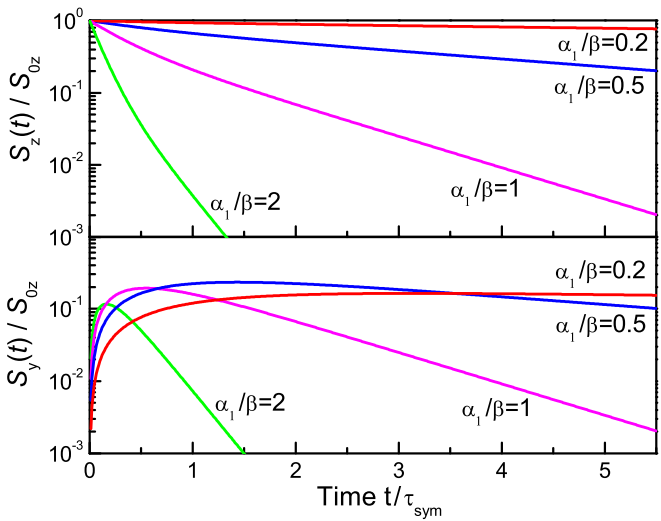}
 \caption{(Color online) Time evolution of the spin components $S_z(t)$ and $S_y(t)$ calculated for different ratios $\alpha_1/\beta$ and the initial spin $\bm{S}_0$ oriented along the QW normal $z$.}
 \label{figure2}
\end{figure}

Figure~\ref{figure2} shows the time dependences of the spin components $S_z(t)$ and $S_y(t)$
given by Eq.~(\ref{S_time}) for different ratios $\alpha_1/\beta$. It is assumed that the spin is initially oriented along the QW normal, $\bm{S}_{0} \parallel z$, and $\alpha_1=\alpha_2$.
The time $t$ is measured here in units of the in-plane relaxation time in symmetric QWs $\tau_{\rm{sym}} = \beta^2 C$. The dependences plotted in logarithmic scale demonstrate that $S_z(t)$ is not described by one exponential function and $S_y$ can be comparable to $S_z$ even in the case of spin pumping along the QW normal. We also note that the sign of $S_y$ is determined by the sign of the product $\alpha_2 \beta$. Therefore, one can control the in-plane spin component by tuning the Rashba field, e.g., by the gate voltage.

\section{Spin relaxation in magnetic field}

Now we consider the effect of an external magnetic field on spin relaxation. The effect is dual. Firstly, the magnetic field causes the precession of nonequilibrium spin resulting in a spin depolarization (Hanle effect). Secondly, the magnetic field leads to the cyclotron motion of free carriers, which changes the direction of the wave vector $\bm{k}$ and thereby suppresses the DP spin relaxation mechanism.~\cite{Ivchenko73,Ivchenko88,Wilamowski04,Glazov04} Both effects can be fruitfully treated in the framework of the spin-density-matrix formalism. In this approach, the electron distribution 
is described by the spin density matrix $\rho_{\bm{k}} = f_{\bm{k}} + \bm{s}_{\bm{k}}\cdot \bm{\sigma}$, where $f_{\bm{k}}$ and $\bm{s}_{\bm{k}}$ are the functions of particle and spin distributions in the $\bm{k}$ space, respectively, and $\bm{\sigma}$ is the vector of the Pauli matrices. The spin distribution $\bm{s}_{\bm{k}}$, within the relaxation time approximation, satisfies the kinetic equation (see, e.g., Refs.~[\onlinecite{Ivchenko73,Glazov04}])
\begin{equation}\label{density_general}
\frac{\partial \bm{s}_{\bm{k}}}{\partial t} +  \bm{s}_{\bm{k}} \times ( \bm{\Omega}_L +\bm{\Omega}_{\bm{k}}) +  \frac{\partial \bm{s}_{\bm{k}}}{\partial k_{\alpha}} \left[ \bm{k} \times \frac{e\bm{B}}{m^* c}  \right]_{\alpha} \hspace{-0.2cm} = \bm{G}_{\bm{k}} - \frac{\bm{s}_{\bm{k}} - \langle \bm{s}_{\bm{k}} \rangle }{\tau_1} \:,
\end{equation}
where $\bm{\Omega}_L$ is the Larmor frequency whose components are given by $\Omega_{L,\alpha}=\mu_B g_{\alpha\beta} B_{\beta}/\hbar$, $\mu_B$ is the Borh magneton, $g_{\alpha\beta}$ is the effective $g$-factor tensor in the absence of cyclotron motion, $e$ is the electron charge, $c$ is the speed of light, and $\bm{G}_{\bm{k}}$ is the spin generation rate into the state $\bm{k}$. We note the $g$-factor can be strongly anisotropic in quantum well structures.~\cite{Ivchenko92,Kalevich92,Salis01} In particular, in (110)-grown QWs of the $C_s$ symmetry the components $g_{xx}$, $g_{yy}$, $g_{zz}$, $g_{yz}$, and $g_{zy}$ are linearly independent and can be non-zero.

The anisotropic part of the spin distribution function $\delta \bm{s}_{\bm{k}} = \bm{s}_{\bm{k}} - \langle \bm{s}_{\bm{k}} \rangle $  is caused by spin-obit splitting of electron states. In the collision-dominated regime, i.e., at $\Omega_{\bm{k}}\tau_1 \ll 1$, it is much less than the isotropic part $\langle \bm{s}_{\bm{k}} \rangle$ and, to the first order in $\Omega_{\bm{k}}\tau_1$, has the form
\begin{equation}\label{delta_s}
\delta \bm{s}_{\bm{k}} = - \frac{\tau_1 \left[ \langle \bm{s}_{\bm{k}} \rangle \times \left( \bm{\Omega}_{\bm{k}} + \omega_c\tau_1 \partial \bm{\Omega}_{\bm{k}} / \partial \varphi_{\bm{k}} \right)\right]}{1+(\omega_c\tau_1)^2} \:.
\end{equation}
Here, $\omega_c = eB_z/(m^*c)$ is the cyclotron frequency, $\varphi_{\bm{k}}$ is the polar angle of the wave vector $\bm{k}$, $\varphi_{\bm{k}} = \arctan(k_y/k_x)$, and $\langle \bm{s}_{\bm{k}} \rangle$ is the quasiequilibrium spin distribution which, in the case of small degree of spin polarization of a two-dimensional electron gas, is given by~\cite{Dyakonov86}  
\begin{equation}\label{s_equil}
\langle \bm{s}_{\bm{k}} \rangle = - \frac{2\pi \hbar^2}{m^* f(0)}  \frac{d
f(\varepsilon_{\bm{k}})}{d\varepsilon_{\bm{k}}} \bm{S} \:.
\end{equation}
Equation~(\ref{delta_s}) is derived taking into account the fact that $\bm{\Omega_{\bm{k}}}$ is linear in $\bm{k}$ and assuming that $\Omega_L \tau_1 \ll 1$ and $\bm{G}_{\bm{k}}$ is an isotropic function of $\bm{k}$.

Finally, substituting expressions given by~(\ref{delta_s}) and~(\ref{s_equil}) for $\delta \bm{s}_{\bm{k}}$ and $\langle \bm{s}_{\bm{k}} \rangle$, respectively, and summing up Eq.~(\ref{density_general}) over the wave vector $\bm{k}$, one obtains the kinetic equation for the total spin density
\begin{equation}\label{S_magn}
\frac{S_{\alpha}(t)}{dt} + [\bm{S}(t) \times \bm{\Omega}'_L ]_{\alpha} = G_{\alpha} - \sum_{\beta} \Gamma'_{\alpha\beta} S_{\beta}(t) \:,
\end{equation}
where $\Gamma_{\alpha\beta}^{\prime}=\Gamma_{\alpha\beta}/[1+(\omega_c\tau_1)^2]$ are components of the spin relaxation rate tensor in the magnetic field, $\bm{\Omega}'_L = \bm{\Omega}_L + \delta \bm{\Omega}_L$, and $\delta \bm{\Omega}_L$ is a contribution to the precession frequency caused by cyclotron motion of free carriers in QWs with spin-orbit splitting (see Ref.~[\onlinecite{Ivchenko73}]),
\begin{equation}\label{Omega_add}
\delta \bm{\Omega}_L = - \frac{\omega_c \tau_1}{1+(\omega_c\tau_1)^2} \int_{0}^{\infty} \frac{\tau_1}{2f(0)} \frac{d f(\varepsilon_{\bm{k}})}{d\varepsilon_{\bm{k}}} \left\langle \bm{\Omega}_{\bm{k}} \times \frac{\partial \bm{\Omega}_{\bm{k}}}{\partial \varphi_{\bm{k}}} \right\rangle d\varepsilon_{\bm{k}} \:.
\end{equation}
In fact, this contribution can be considered as a correction to the effective $g$-factor which depends on the relaxation time $\tau_1$ and the magnetic field strength. We note that the frequency $\delta \bm{\Omega}_L$ is expressed via the same parameters which determine the spin relaxation rate tensor and is comparable to $\Gamma'_{\alpha\beta}$ in the field $\omega_c\tau_1 \sim1$. Taking into account the form of $\bm{\Omega}_{\bm{k}}$ given by Eq.~(\ref{Omega_k}), we obtain non-zero components of $\delta\bm{\Omega}_L$ in (110)-grown QWs
\begin{equation}\label{Omega_add_110}
\delta\Omega_{L,y} = \frac{\alpha_1 \beta \, \omega_c\tau_1}{1+(\omega_c\tau_1)^2} C \:, \;\; \delta\Omega_{L,z} = \frac{\alpha_1 \alpha_2 \, \omega_c\tau_1}{1+(\omega_c\tau_1)^2} C \:.
\end{equation}
\begin{figure}[t]
 \centering\includegraphics[width=0.49\textwidth]{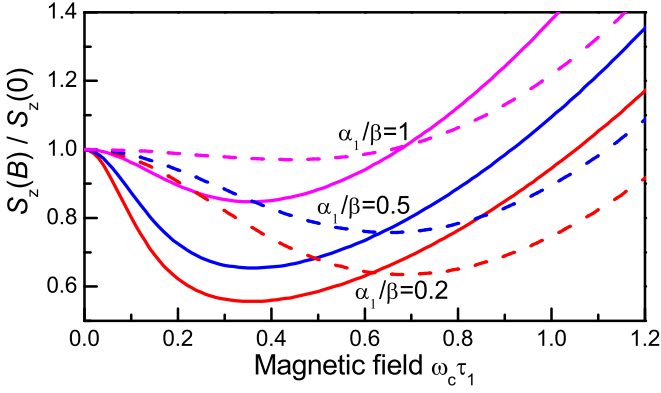}
 \caption{(Color online) Dependences of the electron spin $S_z$ on magnetic field aligned along the QW normal for different ratios $\alpha_1/\beta$. Solid and dashed curves correspond to $\Omega_{L,z} \tau_{\rm{sym}} /\omega_c\tau_1 =5$ and $\Omega_{L,z} \tau_{\rm{sym}} /\omega_c\tau_1 =1$, respectively.}
 \label{figure3}
\end{figure}
Equation~(\ref{S_magn}) describes the spin dynamics in external magnetic field. It yields that
the dependence of the out-of-plane spin component $S_z$, which is usually studied in experiment, on the magnetic field in the case of cw spin pumping along the QW normal has the form
\begin{widetext}
\begin{equation}\label{Sz_B}
S_z = \frac{(\Gamma'_{xx}\Gamma'_{yy}+\Omega_{L,z}^{\prime\,2})G_z} {\Gamma'_{xx} (\Gamma'_{yy}\Gamma'_{zz}-\Gamma_{yz}^{\prime\,2}) + \Gamma'_{xx}\Omega_{L,x}^{\prime \,2} + \Gamma'_{yy}\Omega_{L,y}^{\prime \,2} + \Gamma'_{zz}\Omega_{L,z}^{\prime \,2} + 2\Gamma'_{yz}\Omega_{L,y}^{\prime}\Omega_{L,z}^{\prime} } \:.
\end{equation}
\end{widetext}
Shown in Fig.~\ref{figure3} are the dependences of $S_z$ on the field $\bm{B} \parallel z$ (Faraday geometry) plotted for different ratios $\alpha_1/\beta$ and $\Omega_{L,z} \tau_{\rm{sym}} /\omega_c\tau_1$, $\alpha_1=\alpha_2$ and $g_{yz}=0$. One can see that the magnetic field dependence of $S_z$ can be nonmonotonic even in the Faraday geometry and the effect is more pronounced at $|\Omega_{L,z} \tau_{\rm{sym}} /\omega_c\tau_1| \gg 1$. Such an unusual behavior is attributed to the noncoincidence of the growth direction with a principle axis of the spin relaxation rate tensor, see Fig.~\ref{figure1}.  As it was discussed above, in zero magnetic field, the electron spin initially oriented along $z$ is decomposed into the projections $S_{\tilde{z}}$ and $S_{\tilde{y}}$ which decay at different rates. Since $\gamma_{\tilde{y}}>\gamma_{\tilde{z}}$ for the DP mechanism, the component $S_{\tilde{y}}$ rapidly decays and the spin polarization is determined by the long-life component $S_{\tilde{z}}$. The external magnetic field deflects the electron spin from the slow-relaxation axis $\tilde{z}$ speeding up the spin dephasing. Thus, the Larmor precession leads to the decrease of spin polarization in the magnetic field. At stronger fields, the cyclotron motion suppresses the DP mechanism of spin relaxation, which results in a growth of $S_z$. The interplay of the Larmor precession and the cyclotron motion of free electrons in asymmetric (110) QWs leads to the nonmonotonic dependence of the spin density $S_z$ on magnetic field as shown in Fig.~\ref{figure3}. 

In summary, it is demonstrated that the in-plane and out-of-plane components of electron spin are coupled in quantum wells of low space symmetry. This opens up additional opportunity to manipulate electron spins in low-dimensional semiconductor structures.

\section*{ACKNOWLEDGMENTS}
The author acknowledges fruitful discussions with M.M.~Glazov. This work was supported by the RFBR, program of the Russian Ministry of Education and Science, and the President Grant for young scientists (MD-1717.2009.2).

\newpage

\end{document}